\documentclass[conference]{IEEEtran}

\ifCLASSINFOpdf
  \usepackage[pdftex]{graphicx}
\else
  \usepackage[dvips]{graphicx}
\fi

\usepackage[cmex10]{amsmath}
\usepackage{amssymb}
\usepackage{amsfonts}
\usepackage{bm}
\usepackage{algpseudocode}
\usepackage{algorithm}
\usepackage{array}
\usepackage{url}
\usepackage[noadjust]{cite}
\usepackage{pgfplots}
\pgfplotsset{compat=1.17}
\usepackage{enumerate}
\usepackage{subcaption}
\usepackage{makecell}
\usepackage{booktabs}
\usepackage{xcolor}
\usepackage{acronym}
\usepackage{balance}

\allowdisplaybreaks

\acrodef{led}[LED]{light emitting diode}
\acrodef{irs}[IRS]{intelligent reflecting surface}
\acrodef{nlos}[NLoS]{non-line-of-sight}
\acrodef{los}[LoS]{line-of-sight}
\acrodef{cir}[CIR]{channel impulse response}
\acrodef{cfr}[CFR]{channel frequency response}
\acrodef{snr}[SNR]{signal-to-noise ratio}
\acrodef{pls}[PLS]{physical layer security}
\acrodef{isi}[ISI]{inter-symbol interference}
\acrodef{fov}[FoV]{field of view}
\acrodef{dco-ofdm}[DCO-OFDM]{direct current optical orthogonal frequency division multiplexing}
\acrodef{mrc}[MRC]{maximal-ratio combining}
\acrodef{pd}[PD]{photodetector}
\acrodef{rl}[RL]{reinforcement learning}
\acrodef{mdp}[MDP]{Markov decision process}
\acrodef{ppo}[PPO]{proximal policy optimisation}
\acrodef{csi}[CSI]{channel state information}
\acrodef{rf}[RF]{radio frequency}
\acrodef{6g}[6G]{sixth generation}
\acrodef{lifi}[LiFi]{light fidelity}
\acrodef{iot}[IoT]{internet of things}
\acrodef{ofdm}[OFDM]{orthogonal frequency division multiplexing}
\acrodef{imdd}[IM/DD]{intensity modulation and direct detection}
\acrodef{noma}[NOMA]{non-orthogonal multiple access}
\acrodef{miso}[MISO]{multiple-input single-output}
\acrodef{an}[AN]{artificial noise}
\begin{document}

\title{Non-Uniform Codebook Design for Optical IRS-Assisted VLC Systems}

\author{
Rashid Iqbal\textsuperscript{*},
Dimitrios Bozanis\textsuperscript{\dag},
Dimitrios Tyrovolas\textsuperscript{\dag},
Christos K. Liaskos\textsuperscript{\ddag},\\
Muhammad Ali Imran\textsuperscript{*},
George K. Karagiannidis\textsuperscript{\dag},
Hanaa Abumarshoud\textsuperscript{*}\\[2pt]
\textsuperscript{*}James Watt School of Engineering, University of Glasgow, Glasgow G12 8QQ, U.K.\\
e-mail: r.iqbal.1@research.gla.ac.uk, \{muhammad.imran, hanaa.abumarshoud\}@glasgow.ac.uk\\
\textsuperscript{\dag}Department of Electrical and Computer Engineering, Aristotle University of Thessaloniki, Greece\\
e-mail: \{dimimpoz, tyrovolas, geokarag\}@auth.gr\\
\textsuperscript{\ddag}University of Ioannina and FORTH, Greece\\
e-mail: cliaskos@uoi.gr
}

\maketitle
\thispagestyle{empty}

\begin{abstract}
Optical intelligent reflecting surfaces (OIRS) can improve the coverage of indoor visible light communication (VLC) systems, however, practical deployment requires a finite offline codebook to avoid repeated real-time optimisation of mirror orientations. A uniform codebook with fixed angular steps does not provide uniform coverage on the user plane, because the mapping from steering angles to reflection locations on the user plane is nonlinear. To address this problem, this paper proposes a geometric-optics-based non-uniform codebook design for OIRS-assisted VLC systems. The proposed method constructs an individual codebook for each IRS element according to its geometric position, so that the reflected beams are distributed more uniformly over the user plane. The codebook accuracy is evaluated using the Frobenius norm of the channel error matrix. Simulation results show that the proposed design provides more uniform spatial mapping with fewer codewords than the uniform codebook, and that the sweep-angle resolution has a stronger effect on the codebook accuracy than the tilt-angle resolution.
\end{abstract}

\begin{IEEEkeywords}
Visible light communication, optical intelligent reflecting surfaces, codebook design, geometric optics, beam steering, spatial mapping.
\end{IEEEkeywords}

\section{Introduction}
\IEEEPARstart{V}{isible} light communication (VLC) has emerged as a promising candidate for next-generation wireless systems, driven by the rapid growth in data demand and the increasing congestion of the \ac{rf} spectrum. By exploiting the vast unlicensed optical bandwidth in the visible light range, VLC enables high-rate transmission while leveraging existing \ac{led}-based illumination infrastructure. In addition to its large bandwidth, VLC offers several inherent advantages, including energy efficiency, low electromagnetic interference, enhanced spatial confinement, and accurate localisation, which make it particularly attractive for indoor environments such as offices, hospitals, and industrial settings \cite{haas2016lifi,elgala2011indoor,bozanis_loc}. Despite these benefits, VLC systems still face fundamental limitations in coverage and reliability. Owing to the highly directional nature of optical signals and their strong dependence on \ac{los} propagation, the communication performance is sensitive to blockages, user mobility, and unfavorable geometrical configurations. In practical indoor scenarios, obstacles, user orientation, and shadowing can significantly degrade the received signal strength, leading to coverage holes and unstable performance \cite{ElgalaVLCBlockage}. To maintain connectivity under such conditions, hybrid RF/VLC architectures have been proposed to improve reliability under blocked or degraded optical links \cite{bozanis_hybrid,bozanis_2}. However, these architectures mainly provide connectivity support through the RF link and do not directly reconfigure or improve the optical propagation environment itself.

To directly address the optical propagation limitations of VLC, optical intelligent reflecting surfaces (OIRSs) have recently been proposed as an effective solution for reconfiguring the optical wireless channel. By employing arrays of controllable reflective elements, OIRSs can redirect incident light beams toward desired regions, thereby enhancing coverage, improving signal quality, and enabling flexible environment adaptation \cite{Ahmed_Blockage,abumarshoud2023icc_pls}. Unlike conventional diffuse reflections, OIRSs operate through highly directional specular reflection, where the reflected beam is determined by controllable geometric parameters, typically the tilt and sweep angles of each reflecting element. However, most existing studies assume ideal specular reflection, where the reflected beam is perfectly aligned toward a predefined target location \cite{Ahmed_Blockage,abumarshoud2023icc_pls,Rashid_delay,Rashid_colluding}. Such an assumption implies that the exact reflection angles are known and can be configured with perfect precision, a condition that is difficult to satisfy in practice \cite{SUNCSI}. Even a small deviation in the reflection angles may cause a considerable spatial mismatch between the intended and actual reflection locations on the user plane \cite{bozanis2024}, due to the nonlinear mapping from the angular domain to spatial coordinates. 

To alleviate this angular-sensitivity issue, a codebook-based beam-steering framework has been introduced in \cite{CSIoptics}, where the continuous angular domain is discretised into a finite set of candidate reflection configurations. In this approach, a codeword represents a specific pair of tilt and sweep angles, enabling efficient beam training and robust alignment without requiring exact geometric knowledge. Geometrically, fixing the tilt angle while varying the sweep angle generates reflection points on a ring over the detection plane, whereas varying the tilt angle produces multiple concentric rings with different radii. As a result, the user plane can be systematically covered using a finite number of beam directions. Although codebook-based beam steering improves the practicality of OIRS operation, existing designs are typically constructed using a single reference reflecting element and then reused across the entire OIRS array. This common-codebook assumption overlooks an important geometric feature of multi-element OIRS systems, since different reflecting elements occupy different spatial positions. Consequently, the mapping between steering angles and reflection locations on the user plane becomes element-dependent, and a common codebook may not provide uniform or accurate coverage for all elements.

Motivated by this limitation, this paper develops an element-aware codebook design for multi-element OIRS systems, in which each reflecting element is assigned an individual codebook according to its spatial location. The main novelty of this work lies in replacing the conventional common-codebook assumption with element-specific codebook construction that accounts for the position-dependent angular-to-spatial mapping of each OIRS element. By explicitly incorporating the spatial diversity of the OIRS array, the proposed design enables more accurate and flexible beam steering over the target surface. In addition, the impact of the user-plane height, particularly at \(z = 1\,\text{m}\), is investigated to examine how the distribution of reflection points evolves with the system geometry. This provides practical insight into the interaction between codebook design, OIRS element placement, and spatial coverage in OIRS-assisted VLC systems.

\section{System Model}
\begin{figure}[h]
\centering
\includegraphics[width=0.8\linewidth]{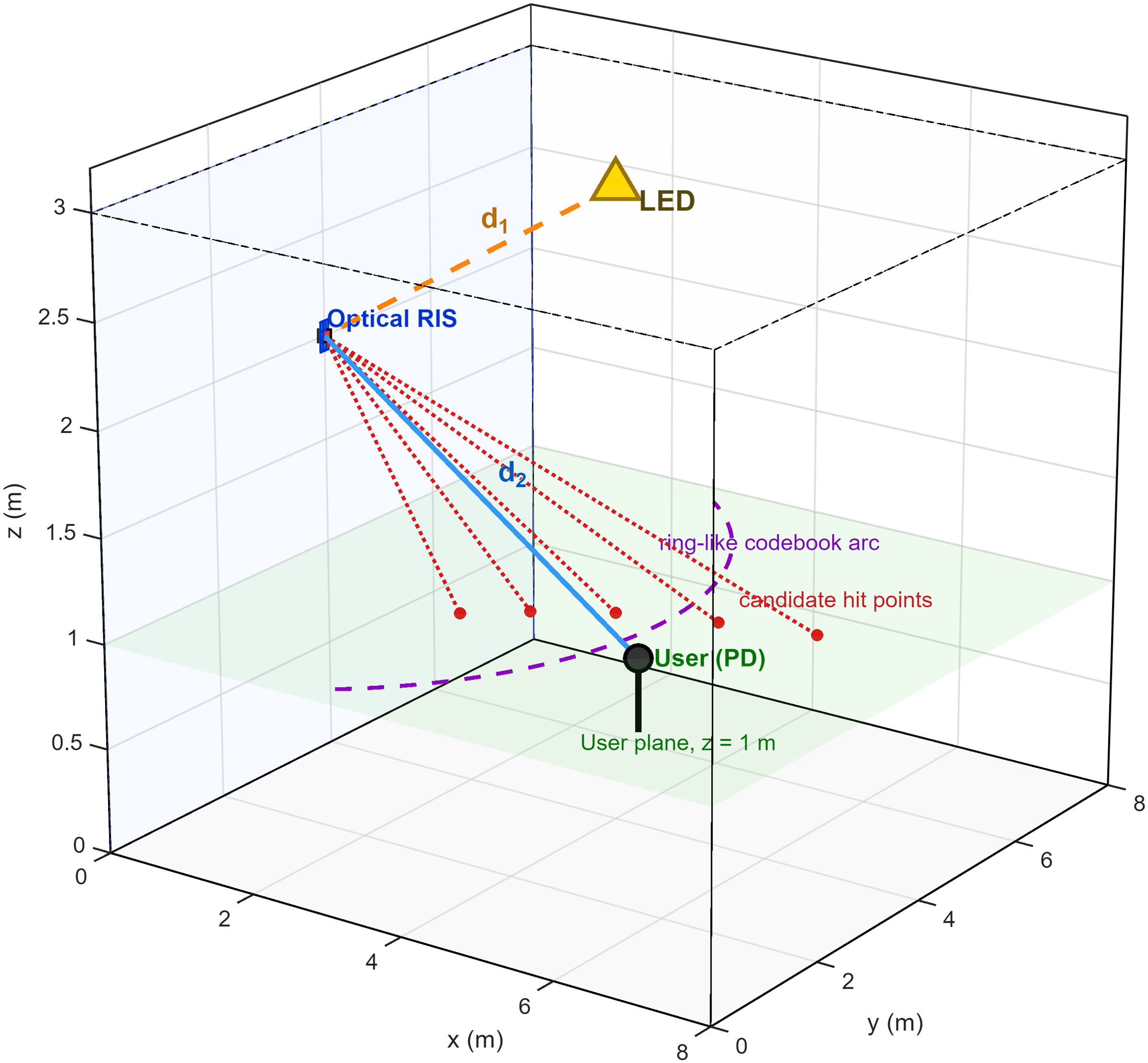}
\caption{Optical IRS System Model with Codebook Design}
\label{SYSTEM MODEL}
\end{figure}

As shown in Fig.~\ref{SYSTEM MODEL}, we consider an indoor VLC system in a room of dimensions $L_x \times L_y \times L_z$ m$^3$. A single LED source is installed on the ceiling at $\mathbf{L}=(x_L,y_L,z_L)$ and radiates downward according to a Lambertian pattern of order $m$. To improve coverage flexibility, an OIRS comprising $N$ independently controllable mirror elements is mounted on a side wall. Specifically, the centre of the $n$-th element, where $n \in \{1,2,\ldots,N\}$, is given by $\mathbf{R}_n=(x_{R_n},y_{R_n},z_{R_n})$. Each element is modelled as a square planar mirror with side length $a$ and reflectivity $\delta$. Moreover, its orientation is controlled through two steering angles, which determine the direction of the reflected optical beam toward the target user plane. The system serves $K$ users distributed on a horizontal plane at height $z_u$. Hence, the position of user $k$ is written as $\mathbf{U}_k=(x_{U_k},y_{U_k},z_u)$, where $z_u < z_{R_n}$ for all reflecting elements. The corresponding photodetector has active area $A$, an upward-facing normal vector $\mathbf{n}_{\mathrm{RX}}=(0,0,1)^T$, and a \ac{fov} half-angle $\Psi_c$. Therefore, only reflected signals arriving within this angular range can be effectively detected.

The received signal at user $k$ consists of a direct
\ac{los} component and a reflected component formed by the superposition of all $N$ OIRS-assisted paths, thus yielding

\begin{equation}
y_k = \left( h_{\mathrm{LoS},k} + \sum_{n=1}^{N}
h_{n,k} \right) s + w_k,
\label{eq:received_signal_new}
\end{equation}

\noindent where $s$ denotes the transmitted optical signal with average power $P_t$, and $w_k \sim \mathcal{N}(0,\sigma_w^2)$ represents additive white Gaussian noise. The direct \ac{los} channel gain follows the standard Lambertian model and is written as \cite{abumarshoud2024irs_vlc_secure}

\begin{small}
\begin{equation}
h_{\mathrm{LoS},k} =
\frac{(m+1)A}{2\pi d_{\mathrm{LoS},k}^2}
\cos^m\!\big(\phi_{\mathrm{LoS},k}\big)
\cos\!\big(\psi_{\mathrm{LoS},k}\big)
\mathbb{I}\!\big(\psi_{\mathrm{LoS},k} \leq \Psi_c\big),
\label{eq:h_los_new}
\end{equation}   
\end{small}

\noindent where $d_{\mathrm{LoS},k} = \|\mathbf{U}_k - \mathbf{L}\|_2$ is the distance between the LED and user $k$, $\phi_{\mathrm{LoS},k}$ is the irradiance angle measured from the LED downward normal $\mathbf{n}_{\mathrm{LED}} = (0,0,-1)^T$, $\psi_{\mathrm{LoS},k}$ is the incidence angle at the user photodetector, and the indicator $\mathbb{I}(\psi_{\mathrm{LoS},k} \leq \Psi_c)$ enforces the receiver \ac{fov} constraint. 

The reflected channel gain from OIRS element $n$ to user $k$ depends on the current mirror orientation and follows the geometric optics model \cite{abumarshoud2024irs_vlc_secure}

\begin{equation}
h_{n,k} =
\frac{(m+1)A\delta}{2\pi(d_{1,n}+d_{2,n,k})^2}
\cos^m\!\big(\phi_n\big)
\cos\!\big(\psi_{n,k}\big)
\mathbb{I}\!\big(\mathbf{Q}_{n,k} \in \mathcal{L}\big),
\label{eq:h_oirs_new}
\end{equation}

\noindent where $d_{1,n} = \|\mathbf{R}_n - \mathbf{L}\|_2$ denotes the distance from the LED to element $n$, $d_{2,n,k} = \|\mathbf{U}_k - \mathbf{R}_n\|_2$ denotes the distance from element $n$ to user $k$, $\phi_n$ is the irradiance angle from the LED toward element $n$, $\psi_{n,k}$ is the incidence angle at user $k$ for the reflected path generated by element $n$, and $\delta$ is the mirror reflectivity, assumed identical for all elements. The indicator $\mathbb{I}(\mathbf{Q}_{n,k} \in \mathcal{L})$ captures the finite size of the LED source by requiring that the backward ray traced from user $k$ through mirror element $n$ intersects the LED aperture region $\mathcal{L}$, namely $\|\mathbf{Q}_{n,k} - \mathbf{L}\|_2 \leq r_{\mathrm{LED}}$, where $r_{\mathrm{LED}}$ is the LED aperture radius. This transmitter-side constraint is fundamentally different from the \ac{fov} constraint in \eqref{eq:h_los_new}, since the latter acts at the receiver side.

The aggregate reflected gain and the overall optical channel gain at user $k$ are therefore given by

\begin{equation}
h_{\mathrm{OIRS},k} = \sum_{n=1}^{N} h_{n,k},
\qquad
h_{\mathrm{tot},k} = h_{\mathrm{LoS},k} + h_{\mathrm{OIRS},k},
\label{eq:h_total_new}
\end{equation}

\noindent and the corresponding electrical signal-to-noise ratio under intensity modulation and direct detection is

\begin{equation}
\mathrm{SNR}_k =
\frac{h_{\mathrm{tot},k}^2 P_t^2}{\sigma_w^2}.
\label{eq:snr_new}
\end{equation}

\noindent The quadratic dependence on $h_{\mathrm{tot},k}$ appears because the received optical power is first converted linearly to electrical current, after which the electrical SNR becomes proportional to the square of the optical channel gain.

The central challenge in such a system is that the mirror orientation of each element cannot be continuously recomputed for every possible user location during real-time operation. Therefore, each element must rely on a finite set of precomputed steering configurations, namely a codebook $\mathcal{C}_n$, designed offline for element $n$. During operation, element $n$ selects, for user $k$, the codeword whose corresponding hit point on the detection plane is closest to the user location, i.e.,

\begin{equation}
\mathbf{c}_{n,k}^{\star}
=
\arg\min_{\mathbf{c}\in\mathcal{C}_n}
\left\|\mathbf{P}_n(\mathbf{c}) - \mathbf{U}_k\right\|_2,
\label{eq:codeword_select_new}
\end{equation}

\noindent where $\mathbf{P}_n(\mathbf{c})$ denotes the point reached on the detection plane by the reflected beam from element $n$ under codeword $\mathbf{c}$. This selection induces a codebook-based gain matrix $\mathbf{G} \in \mathbb{R}_{+}^{N \times K}$ whose entries are

\begin{equation}
g_{n,k}
=
h_{n,k}\!\left(\mathbf{R}_n,\mathbf{U}_k \mid \mathbf{c}_{n,k}^{\star}\right),
\label{eq:G_matrix_new}
\end{equation}

\noindent  which denotes the reflected channel gain achieved by element $n$ for user $k$ under the selected codeword $\mathbf{c}_{n,k}^{\star}$

\section{GO-Based Non-Uniform OIRS Codebook Design}
\label{sec:codebook}

The steering state of each OIRS mirror element is fully characterised by two angles, namely the tilt angle $\vartheta \in [-\pi/2,\pi/2)$ and the sweep angle $\varphi \in [-\pi/2,\pi/2)$. The tilt angle mainly controls how far the reflected beam reaches on the detection plane, whereas the sweep angle controls the lateral position of the reflected beam. Accordingly, the unit normal vector of the mirror can be written as

\begin{equation}
\widehat{\mathbf{n}}(\vartheta,\varphi) =
(\cos\vartheta \sin\varphi, \cos\vartheta \cos\varphi, -\sin\vartheta)
\label{eq:normal_new}
\end{equation}

For an incident optical ray impinging on the mirror, the outgoing reflected direction is determined by the law of specular reflection, namely

\begin{equation}
\mathbf{v}_{\mathrm{out}} =
\mathbf{v}_{\mathrm{in}} -
2\big(\mathbf{v}_{\mathrm{in}} \cdot
\widehat{\mathbf{n}}(\vartheta,\varphi)\big)
\widehat{\mathbf{n}}(\vartheta,\varphi),
\label{eq:reflection_new}
\end{equation}

\noindent where $\mathbf{v}_{\mathrm{in}} = (\mathbf{R}_n - \mathbf{L})/\|\mathbf{R}_n - \mathbf{L}\|_2$ is the incident unit vector from the LED to element $n$. Once the reflected direction is obtained, the corresponding reflection points on the user plane of height $z_u$ is determined through ray tracing as

\begin{equation}
\mathbf{P}_n(\vartheta,\varphi) =
\mathbf{R}_n + \tau \mathbf{v}_{\mathrm{out}},
\qquad
\tau = \frac{z_u - z_{R_n}}{v_{\mathrm{out},z}},
\label{eq:phit_new}
\end{equation}

\noindent where $v_{\mathrm{out},z}$ is the $z$-component of $\mathbf{v}_{\mathrm{out}}$. The condition $v_{\mathrm{out},z} < 0$ guarantees that the reflected ray propagates downward and physically intersects the detection plane.

This geometric construction fully specifies the codeword selection rule in \eqref{eq:codeword_select_new}, since each pair $(\vartheta,\varphi)$ maps to a unique point on the detection plane. If the tilt angle is fixed and the sweep angle is varied continuously, the reflected beam traces a circular arc on the detection plane, which is hereafter referred to as a ring. Successive values of the tilt angle generate multiple
rings with different radii. The radius of a ring is governed by the tilt angle through a nonlinear geometric relation. To express this relation compactly, define

\begin{equation}
\beta_n =
\arccos\!\left(
\mathbf{e}_3^T
\frac{\mathbf{R}_n - \mathbf{L}}
{\|\mathbf{R}_n - \mathbf{L}\|_2}
\right),
\label{eq:beta_new}
\end{equation}

\noindent where $\mathbf{e}_3 = (0,0,1)^T$ denotes the vertical axis. Owing to the law of reflection, rotating the mirror by $\vartheta$ rotates the reflected beam by $2\vartheta$, hence the ring radius is proportional to $\tan(\beta_n+2\vartheta)$.

A straightforward uniform codebook samples the angular domain using fixed steps in both steering variables. However, such a design does not yield uniform spatial coverage on the detection plane. First, equal increments in $\vartheta$ do not generate equal radial gaps between successive rings because of the nonlinear tangent relation. Second, equal increments in $\varphi$ lead to unequal arc lengths on rings of different radii. Consequently, many codewords become spatially redundant, particularly near the inner region, whereas outer regions remain more sparsely covered.

To overcome this mismatch, we construct a geometric-optics-based non-uniform codebook for each element individually. The design is based on two principles, equal radial spacing between neighbouring rings on the detection plane, and approximately equal arc length between adjacent codewords on each ring. In this way, the codebook points become more uniformly distributed over the user plane, even though the steering angles themselves are sampled non-uniformly.

The construction begins by computing the reference sweep angle for element $n$ as

\begin{equation}
\varphi_n^{\mathrm{c}} =
\mathrm{atan2}\!\left(
\mathbf{e}_2^T \mathbf{v}_{\mathrm{in}},
\mathbf{e}_1^T \mathbf{v}_{\mathrm{in}}
\right),
\label{eq:varphi_c_new}
\end{equation}

\noindent where $\mathbf{e}_1$ and $\mathbf{e}_2$ are the unit vectors along the $x$ and $y$ axes, respectively. This reference angle identifies the central sweep direction for the mirror and serves as the anchor around which the codebook is constructed.

The first tilt codeword for element $n$ is then chosen as the mirror orientation that steers the reflected beam vertically downward toward the point directly below the element on the detection plane, namely  \cite{SUNCSI}

\begin{equation}
\vartheta_{n,1} =
-\arccos\!\left(
\frac{
\left(
\frac{\mathbf{R}_n - \mathbf{L}}
{\|\mathbf{R}_n - \mathbf{L}\|_2}
+ \mathbf{e}_3
\right)^T
\widehat{\mathbf{n}}(0,\varphi_n^{\mathrm{c}})
}
{
\left\|
\frac{\mathbf{R}_n - \mathbf{L}}
{\|\mathbf{R}_n - \mathbf{L}\|_2}
+ \mathbf{e}_3
\right\|_2
}
\right).
\label{eq:vartheta1_new}
\end{equation}

Starting from this initial codeword, the second tilt level is initialised as

\begin{equation}
\vartheta_{n,2} = \vartheta_{n,1} - \Delta_{\vartheta},
\label{eq:vartheta2_new}
\end{equation}

\noindent where $\Delta_{\vartheta}$ is a design parameter that controls the initial ring density. To maintain equal radial spacing on the detection plane, the subsequent tilt codewords are generated such that the transformed quantities $\tan(\beta_n+2\vartheta_{n,i})$ form an arithmetic progression, i.e.,

\begin{equation}
\tan(\beta_n + 2\vartheta_{n,i+1}) + \tan(\beta_n + 2\vartheta_{n,i-1})
= 2\tan(\beta_n + 2\vartheta_{n,i}),
\label{eq:recursion_new}
\end{equation}

\noindent from which the next tilt codeword follows as

\begin{small}
\begin{equation}
\vartheta_{n,i+1}
=
\frac{1}{2}
\left[
\arctan\!\Big(
2\tan(\beta_n + 2\vartheta_{n,i}) -
\tan(\beta_n + 2\vartheta_{n,i-1})
\Big)
-
\beta_n
\right].
\label{eq:vartheta_next_new}
\end{equation}  
\end{small}

Although the resulting tilt angles are not equally spaced in the angular domain, this non-uniformity is intentional, since it compensates for the tangent nonlinearity and makes the corresponding ring radii more uniformly spaced on the detection plane.

Once the $i$-th ring is fixed through $\vartheta_{n,i}$, the associated sweep codewords are generated so that the arc length between neighbouring codewords remains approximately constant. This is achieved by reducing the angular sweep spacing with the ring index, namely

\begin{equation}
\varphi_{n,i,\ell}
=
\varphi_n^{\mathrm{c}}
\pm
\ell \frac{\Delta_{\varphi}}{i},
\qquad
\ell = 1,2,3,\ldots
\label{eq:varphi_new}
\end{equation}

\noindent until the generated sweep angles fall outside the feasible angular range. Here $\Delta_{\varphi}$ is the second design parameter controlling the base sweep resolution. Since outer rings have larger radii, they require more codewords to preserve the same spatial arc length, which is precisely achieved by the factor $1/i$ in \eqref{eq:varphi_new}. Therefore, the complete codebook of element $n$ is constructed from the set of tilt levels $\{\vartheta_{n,i}\}$ and the corresponding sweep sets $\{\varphi_{n,i,\ell}\}$. The complete procedure for constructing the proposed element-wise non-uniform codebook is summarised in Algorithm~\ref{alg:codebook_new}.

The quality of the resulting codebooks is evaluated through the Frobenius norm of the channel error matrix over the sampled user plane. Specifically, the mismatch between the ideal continuous steering and the codebook-based steering is measured as

\begin{equation}
\|\Delta \mathbf{H}\|_F
=
\sqrt{
\sum_{n=1}^{N}
\sum_{k}
\left(
h_{\mathrm{ideal},n}(\mathbf{U}_k)
-
h_{\mathrm{cb},n}(\mathbf{U}_k)
\right)^2
},
\label{eq:frobenius_new}
\end{equation}

\noindent where $h_{\mathrm{ideal},n}(\mathbf{U}_k)$ and $h_{\mathrm{cb},n}(\mathbf{U}_k)$ denote the channel gains obtained from element $n$ under ideal continuous steering and codebook-based steering, respectively, at user position $\mathbf{U}_k$. A smaller value of $\|\Delta \mathbf{H}\|_F$ indicates that the discrete codebook more accurately approximates the ideal beam alignment over all elements and user locations.

\begin{algorithm}[h]
\vspace{0.06in}
\caption{Element-Wise GO-Based Non-Uniform OIRS Codebook Generation}
\label{alg:codebook_new}
\begin{algorithmic}[1]
\Require Element position $\mathbf{R}_n$, LED position $\mathbf{L}$, design parameters $\Delta_{\vartheta}$, $\Delta_{\varphi}$
\Ensure Codebook $\mathcal{C}_n$
\State Compute the incident direction
$\mathbf{v}_{\mathrm{in}} =
(\mathbf{R}_n - \mathbf{L})/\|\mathbf{R}_n - \mathbf{L}\|_2$
\State Compute $\beta_n$ from \eqref{eq:beta_new}
\State Compute the reference sweep angle $\varphi_n^{\mathrm{c}}$ from \eqref{eq:varphi_c_new}
\State Compute the first tilt codeword $\vartheta_{n,1}$ from \eqref{eq:vartheta1_new}
\State Initialise the second tilt codeword as $\vartheta_{n,2} = \vartheta_{n,1} - \Delta_{\vartheta}$
\State Initialise $\mathcal{C}_n \leftarrow \emptyset$, set $i \leftarrow 1$
\Repeat
    \State Generate sweep angles for ring $i$ as
    $\varphi_{n,i,\ell} =
    \varphi_n^{\mathrm{c}} \pm \ell \Delta_{\varphi}/i$
    for all valid $\ell$
    \State Add all valid pairs $(\vartheta_{n,i}, \varphi_{n,i,\ell})$ to $\mathcal{C}_n$
    \State Compute next tilt angle $\vartheta_{n,i+1}$ from \eqref{eq:vartheta_next_new}
    \State Update $i \leftarrow i+1$
\Until{the reflected points fall outside the user plane}
\end{algorithmic}
\end{algorithm}

\section{Simulation Results}
This section highlights the importance of designing a non-uniform codebook for an OIRS. In practical systems, the mirror orientations cannot be continuously updated for every user position, and therefore a finite offline codebook must be used. A uniform codebook that samples the angular domain with fixed step sizes does not provide uniform coverage on the user plane due to the nonlinear mapping between angles and spatial locations. This results in a high concentration of beams near the IRS and sparse coverage farther away, leading to inefficient use of codewords. In contrast, the proposed non-uniform codebook is designed based on geometric optics to achieve a more uniform spatial distribution of reflected beams. This allows the system to approximate the ideal beam steering more accurately, while maintaining low computational complexity during real-time operation. Unless otherwise stated, the simulations are conducted in an indoor room of size $8 \times 8 \times 3~\mathrm{m}^3$, and the main system parameters are summarised in Table~\ref{tab:sim_params}.

\begin{table}[t]
\vspace{0.06in}
\centering
\caption{Simulation parameters \cite{Rashid_colluding}.}
\label{tab:sim_params}
\begin{tabular}{ll}
\hline
Parameter & Value \\
\hline
Room size & $8 \times 8 \times 3~\mathrm{m}^3$ \\
LED position & $(4,4,3)$ \\
OIRS centre position & $(0,4,2)$ \\
OIRS size & $3 \times 3$ elements \\
Element spacing & $0.09$ m \\
User plane height & $z_u = 1$ m \\
Lambertian order & $m = 1$ \\
PD area & $A = 10^{-4}\,\mathrm{m}^2$ \\
Mirror reflectivity & $\delta = 1$ \\
Field of view & $\Psi_c = 90^\circ$ \\
\hline
\end{tabular}
\end{table}

\begin{figure}[h]
\centering
\includegraphics[width=0.8\linewidth]{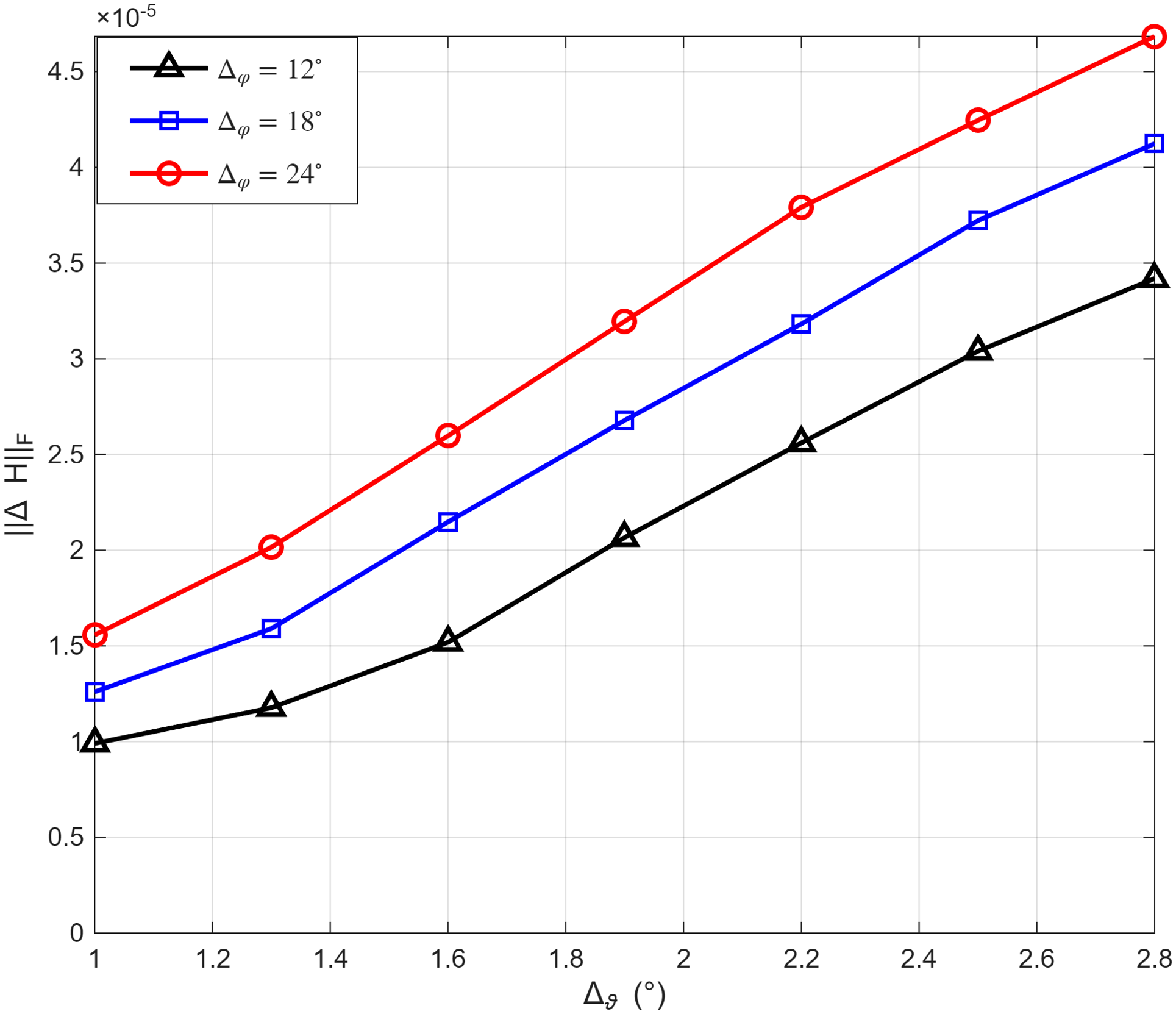}
\caption{Frobenius norm $\|\Delta \mathbf{H}\|_F$ versus $\Delta_{\vartheta}$.}
\label{fig:omega_new}
\end{figure}

\begin{figure}[h]
\centering
\includegraphics[width=0.8\linewidth]{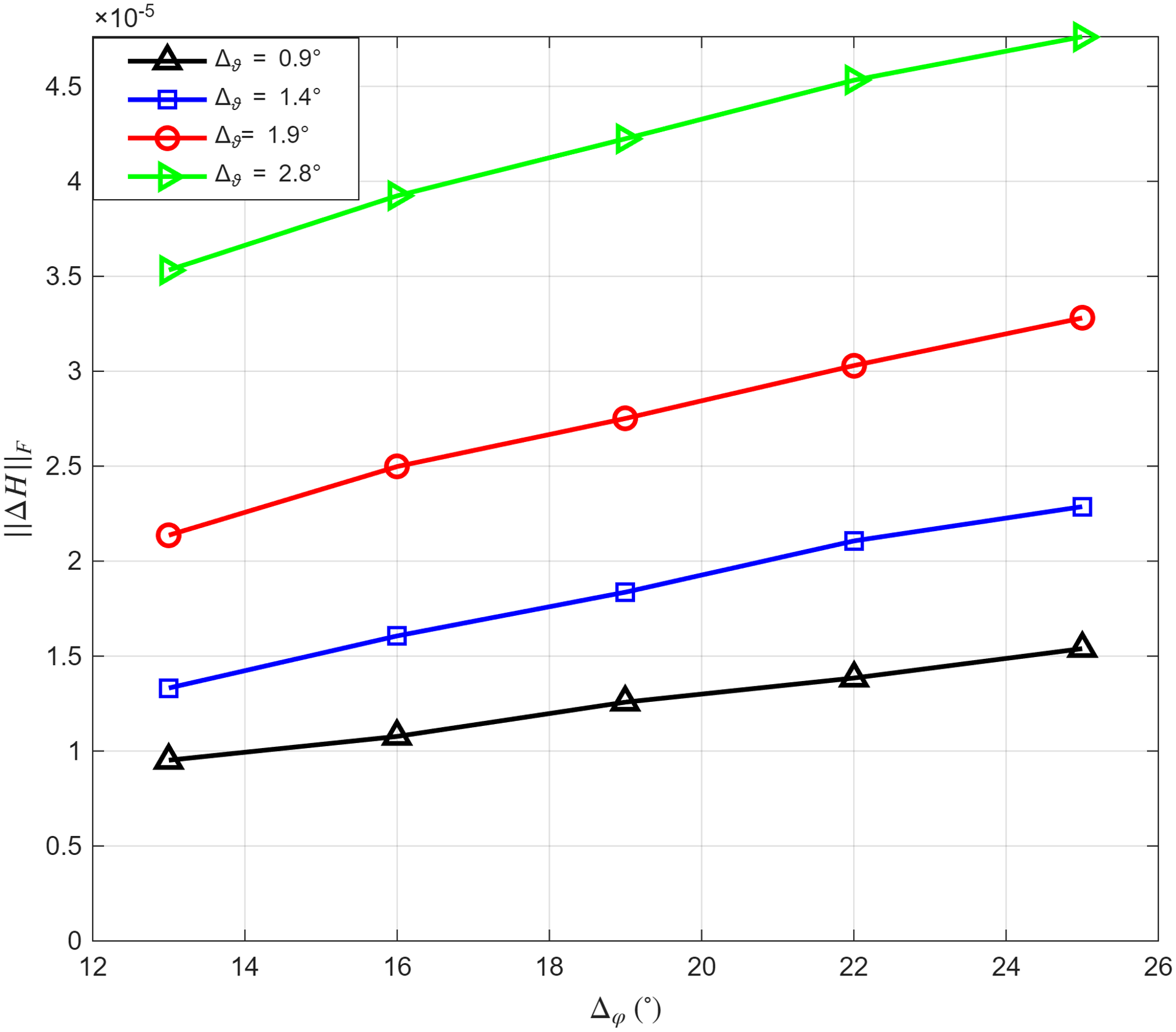}
\caption{Frobenius norm $\|\Delta \mathbf{H}\|_F$ versus $\Delta_{\varphi}$.}
\label{fig:gamma_new}
\end{figure}

Fig.~\ref{fig:omega_new} and Fig.~\ref{fig:gamma_new} together show how the step sizes $\Delta_{\vartheta}$ and $\Delta_{\varphi}$ affect the accuracy of the proposed codebook. It can be observed that the channel error increases when either of the step sizes becomes larger, since the codebook becomes less dense and the beam alignment becomes less accurate. However, the effect of $\Delta_{\vartheta}$ is relatively small, especially when it is already chosen to be sufficiently fine, as it mainly controls the radial distribution of the rings. In contrast, $\Delta_{\varphi}$ has a much stronger impact on performance because it determines how closely the codewords are spaced along each ring. A larger $\Delta_{\varphi}$ results in fewer points along the arc, which leads to a higher mismatch between the desired and actual reflection directions. Therefore, the results indicate that improving the angular resolution along the rings is more important for reducing the error, while the tilt resolution provides additional refinement once a reasonable radial coverage is achieved.

\begin{figure}[h]
\centering
\includegraphics[width=0.85\linewidth]{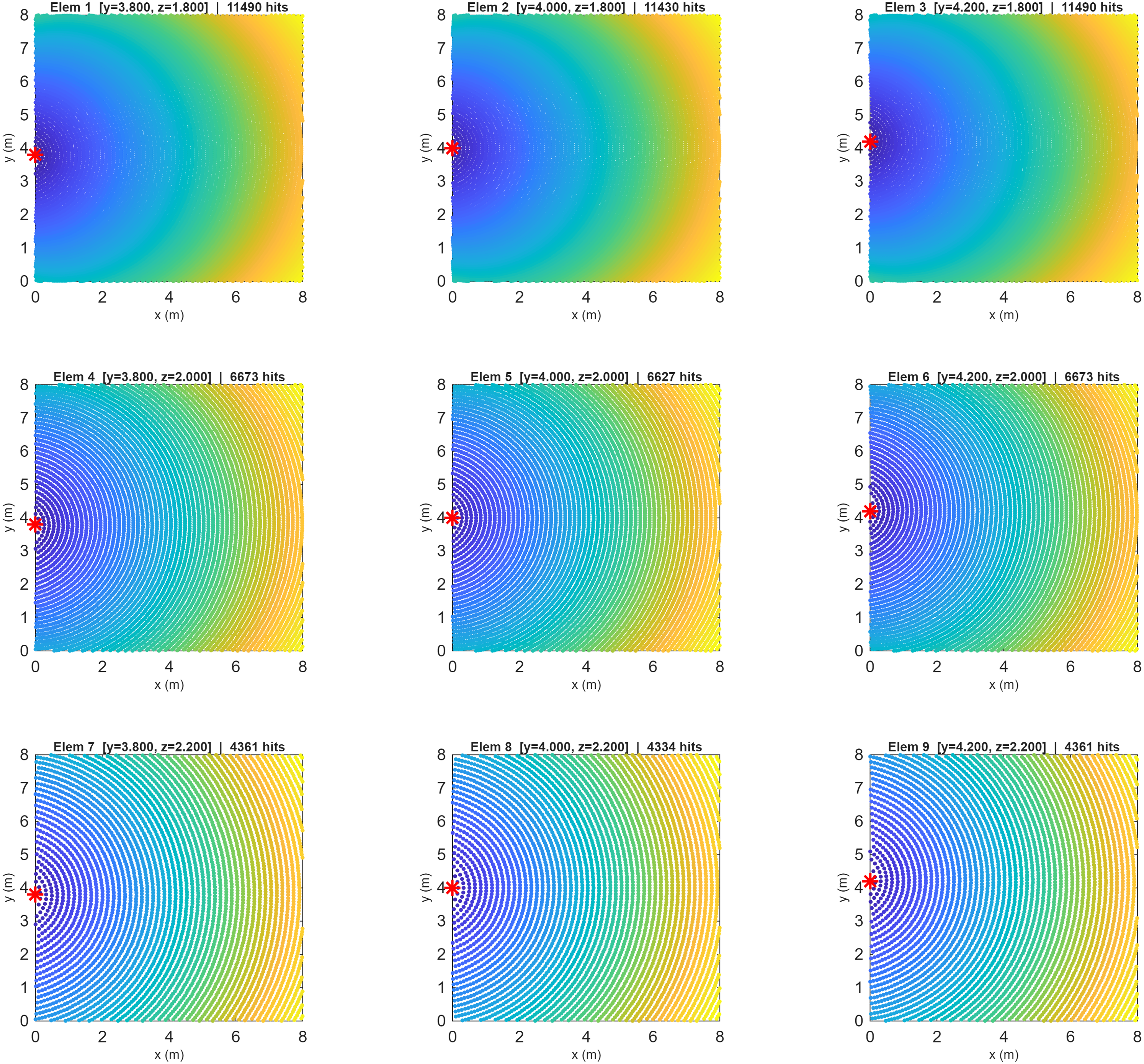}
\caption{Per-element non-uniform codebook.}
\label{fig:radius_new}
\end{figure}

Fig.~\ref{fig:radius_new} illustrates the element-dependent spatial mapping of the proposed non-uniform codebook for a \(3 \times 3\) OIRS. Since each reflecting element occupies a different position on the OIRS plane, the incident direction, reference sweep angle, and generated steering codewords are computed separately for each element. Consequently, the reflected beam-hit points on the user plane are shifted differently for different elements, even though the same codebook-generation principle is applied. This confirms that a single reference-element codebook cannot accurately represent the spatial mapping of all OIRS elements. Therefore, the proposed element-wise design captures the geometry-dependent behaviour of multi-element OIRS beam steering and provides a more practical basis for constructing spatially balanced codebooks.

\begin{figure}[h]
\centering
\includegraphics[width=0.85\linewidth]{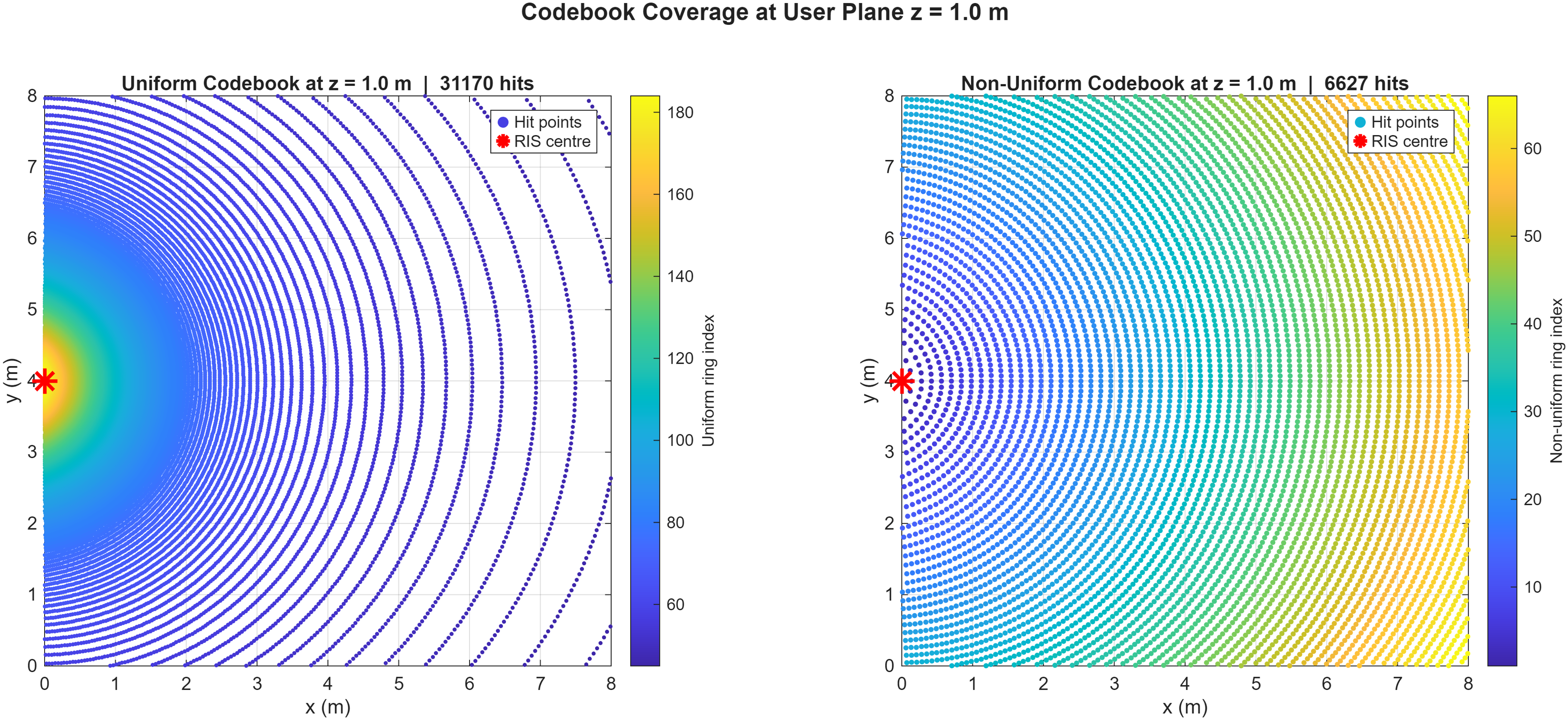}
\caption{Uniform versus non-uniform codebook.}
\label{fig:radius_new1}
\end{figure}

Fig.~\ref{fig:radius_new1} compares the spatial coverage of the uniform and non-uniform codebooks at the user plane. In the uniform case, where fixed step sizes are used for all directions, it can be observed that the reflected points are highly concentrated near the IRS, resulting in a very dense mapping in the nearby region. However, as the distance from the IRS increases, the spacing between the points becomes larger, leading to poor coverage and visible gaps in the farther areas. In addition, the total number of reflection points is very high, which increases the complexity without providing uniform coverage. In contrast, the proposed non-uniform codebook uses adaptive step sizes, which results in a more uniform distribution of reflection points across the entire user plane. The mapping does not suffer from large gaps, and the coverage remains consistent even at larger distances from the IRS. At the same time, the number of reflection points is significantly reduced compared to the uniform case, which improves efficiency. Therefore, the non-uniform design achieves a better trade-off between coverage and complexity, providing uniform spatial mapping with fewer codewords.

\begin{figure}[h]
\centering
\includegraphics[width=0.8\linewidth]{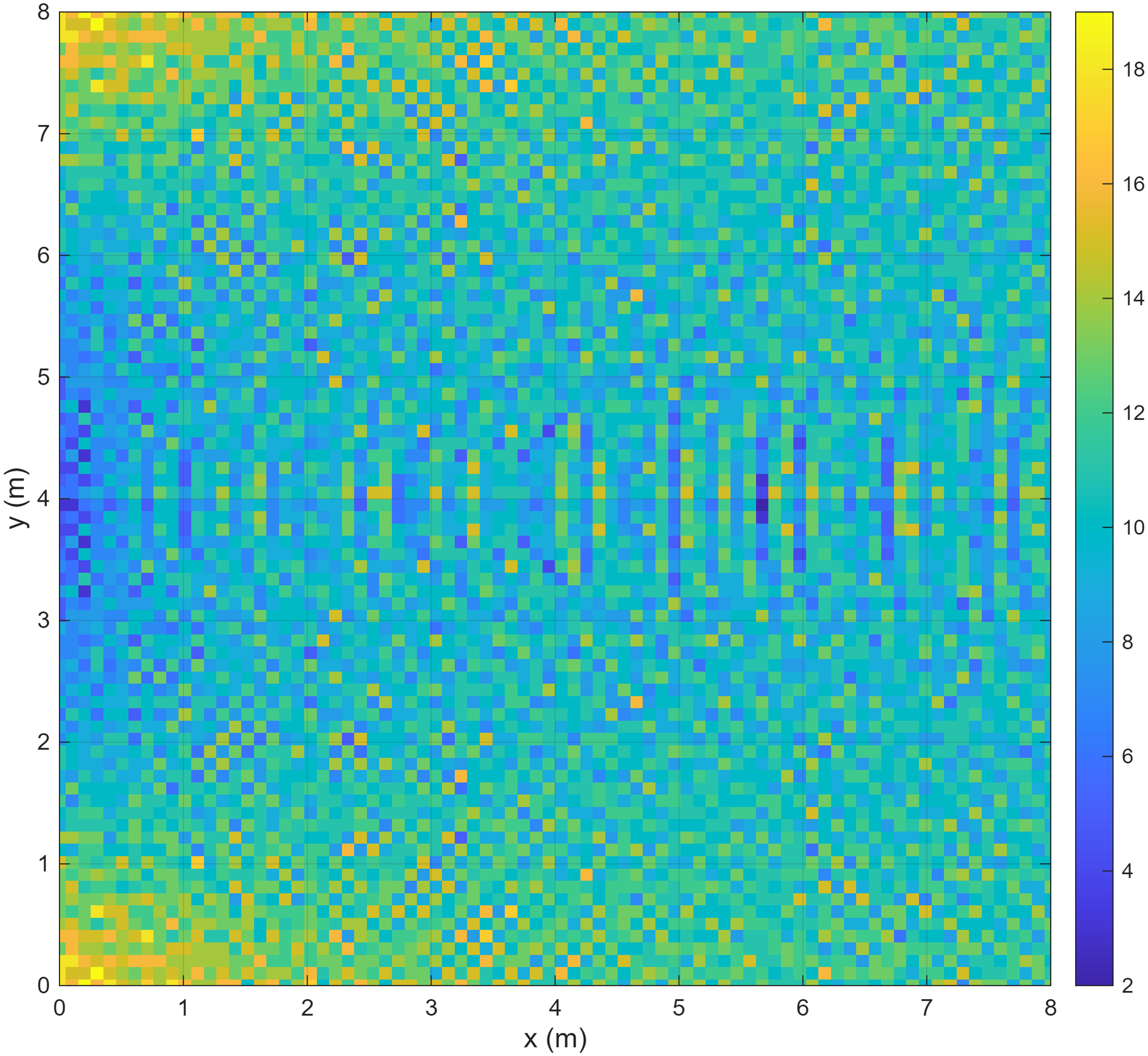}
\caption{Aggregated Hit Density at z = 1.0 m.}
\label{fig:radius_new2}
\end{figure}

Fig.~\ref{fig:radius_new2} shows the aggregated hit density at the user plane for the proposed $3 \times 3$ OIRS system, where the contributions from all IRS elements are combined. Each point in the figure represents how many times a location on the user plane is reached by the reflected beams from different codewords and different IRS elements. It can be observed that the hit density is distributed over the entire area without large empty regions, which indicates that the proposed non-uniform codebook provides good spatial coverage when all elements operate together. In addition, although some variations in density can be seen due to the different geometric positions of the IRS elements, the overall distribution remains relatively balanced across the plane. This confirms that the geometry-aware design of the non-uniform codebook allows different elements to complement each other, reducing coverage gaps. Therefore, the aggregated result demonstrates that the proposed approach achieves effective and uniform coverage of the user plane with a limited number of codewords, making it suitable for practical multi-element OIRS deployments.

% ============================================================

\section{Conclusion}

This paper proposes a geometric-optics based non-uniform codebook design for OIRS-assisted VLC systems. Unlike a uniform codebook, the proposed design accounts for the nonlinear mapping between steering angles and beam reflection points on the user plane, which leads to more balanced spatial coverage. The method has been further extended to a multi-element OIRS by constructing a separate codebook for each RIS element according to its geometric position. The results show that the proposed codebook reduces the channel mismatch error and achieves more uniform coverage with fewer codewords than the uniform design. The results also indicate that the sweep-angle resolution has a larger effect on the codebook accuracy than the tilt-angle resolution.

\section*{Acknowledgement}
This work was supported by the EPSRC project UKRI916 PerceptiFi: Integrated Light Sensing and Communications for Perceptive LiFi.
\balance
% \bibliographystyle{IEEEtran}
% \bibliography{References}

\begin{thebibliography}{10}
\providecommand{\url}[1]{#1}
\csname url@samestyle\endcsname
\providecommand{\newblock}{\relax}
\providecommand{\bibinfo}[2]{#2}
\providecommand{\BIBentrySTDinterwordspacing}{\spaceskip=0pt\relax}
\providecommand{\BIBentryALTinterwordstretchfactor}{4}
\providecommand{\BIBentryALTinterwordspacing}{\spaceskip=\fontdimen2\font plus
\BIBentryALTinterwordstretchfactor\fontdimen3\font minus \fontdimen4\font\relax}
\providecommand{\BIBforeignlanguage}[2]{{%
\expandafter\ifx\csname l@#1\endcsname\relax
\typeout{** WARNING: IEEEtran.bst: No hyphenation pattern has been}%
\typeout{** loaded for the language `#1'. Using the pattern for}%
\typeout{** the default language instead.}%
\else
\language=\csname l@#1\endcsname
\fi
#2}}
\providecommand{\BIBdecl}{\relax}
\BIBdecl

\bibitem{haas2016lifi}
H.~Haas, L.~Yin, Y.~Wang, and C.~Chen, ``{What is {LiFi}?},'' \emph{J. Lightw.
  Technol.}, vol.~34, no.~6, pp. 1533--1544, Mar. 15 2016.

\bibitem{elgala2011indoor}
H.~Elgala, R.~Mesleh, and H.~Haas, ``{Indoor optical wireless communication:
  potential and state-of-the-art},'' \emph{IEEE Commun. Mag.}, vol.~49, no.~9,
  pp. 56--62, 2011.

\bibitem{bozanis_loc}
D.~Bozanis, D.~Tyrovolas, V.~K. Papanikolaou, S.~A. Tegos, P.~D.
  Diamantoulakis, C.~K. Liaskos, R.~Schober, and G.~K. Karagiannidis,
  ``{Closed-form location and orientation estimation in optical wireless
  systems},'' in \emph{IEEE WCNC 2025 - IEEE Wireless Commun. Netw. Conf.},
  2025, pp. 1--6.

\bibitem{ElgalaVLCBlockage}
H.~Elgala, R.~Mesleh, and H.~Haas, ``{Indoor optical wireless communication:
  potential and state-of-the-art},'' \emph{IEEE Commun. Mag.}, vol.~49, no.~9,
  pp. 56--62, 2011.

\bibitem{bozanis_hybrid}
D.~Bozanis, V.~K. Papanikolaou, A.~A. Dowhuszko, K.~G. Rallis, P.~D.
  Diamantoulakis, J.~H{\"a}m{\"a}l{\"a}inen, and G.~K. Karagiannidis,
  ``{Optimal Aggregation of RF and VLC Bands for Beyond 5G Mobile Services},''
  in \emph{2022 18th International Conference on Wireless and Mobile Computing,
  Networking and Communications (WiMob)}, 2022, pp. 75--80.

\bibitem{bozanis_2}
D.~Bozanis, N.~A. Mitsiou, S.~A. Tegos, P.~D. Diamantoulakis, G.~K.
  Karagiannidis, and V.~K. Papanikolaou, ``{On the Beamforming Design of
  Cross-Band OWC/RF Cell-Free MIMO},'' \emph{IEEE Commun. Lett.}, pp. 1--1,
  2025.

\bibitem{Ahmed_Blockage}
A.~R. Hussen, R.~Iqbal, A.~Zoha, M.~A. Imran, and H.~Abumarshoud, ``{Sum rate
  maximisation for irs-assisted vlc using reinforcement learning},'' in
  \emph{2024 IEEE Middle East Conference on Communications and Networking
  (MECOM)}, 2024, pp. 464--469.

\bibitem{abumarshoud2023icc_pls}
H.~Abumarshoud, C.~Chen, I.~Tavakkolnia, H.~Haas, and M.~A. Imran,
  ``{Intelligent Reflecting Surfaces for Enhanced Physical Layer Security in
  NOMA VLC Systems},'' in \emph{ICC 2023 - IEEE Int. Conf. Commun.}, 2023, pp.
  3284--3289.

\bibitem{Rashid_delay}
R.~Iqbal, M.~Biagi, A.~Zoha, M.~A. Imran, and H.~Abumarshoud, ``{Leveraging
  IRS Induced Time Delay for Enhanced Physical Layer Security in VLC Systems},''
  \emph{IEEE Wireless Commun. Lett.}, vol.~13, no.~11, pp. 3147--3151, 2024.

\bibitem{Rashid_colluding}
R.~Iqbal, A.~Zoha, S.~Ikki, M.~A. Imran, and H.~Abumarshoud, ``{Enhancing pls
  of indoor irs-vlc systems for colluding and non-colluding eavesdroppers},''
  \emph{IEEE Open J. Commun. Soc.}, vol.~7, pp. 2765--2776, 2026.

\bibitem{SUNCSI}
S.~Sun, F.~Yang, W.~Mei, J.~Song, Z.~Han, and R.~Zhang, ``{Channel estimation
  for optical intelligent reflecting surface-assisted vlc system: A joint
  space-time sampling approach},'' \emph{IEEE J. Sel. Areas Commun.}, vol.~43,
  no.~3, pp. 867--882, 2025.

\bibitem{bozanis2024}
D.~Bozanis, D.~Tyrovolas, V.~K. Papanikolaou, S.~A. Tegos, P.~D.
  Diamantoulakis, C.~K. Liaskos, R.~Schober, and G.~K. Karagiannidis,
  ``{Location-Driven Programmable Wireless Environments through Light-emitting
  RIS (LeRIS)},'' 2024. [Online]. Available: https://arxiv.org/abs/2412.04989

\bibitem{CSIoptics}
S.~Sun, F.~Yang, W.~Mei, J.~Song, Z.~Han, and R.~Zhang, ``{Optical irs for
  visible light communication: From optics model to association model},''
  \emph{IEEE Wireless Commun.}, vol.~32, no.~4, pp. 162--169, 2025.

\bibitem{abumarshoud2024irs_vlc_secure}
H.~Abumarshoud and M.~Biagi, ``{Intelligent Reflecting Surface-Aided Visible
  Light Communications for Granting Indoor Secrecy},'' in \emph{ICC 2024 -
  IEEE Int. Conf. Commun.}, 2024, pp. 3701--3706.

\end{thebibliography}

\end{document}